\documentclass[sigconf, screen]{acmart}
\AtBeginDocument{%
  \providecommand\BibTeX{{%
    \normalfont B\kern-0.5em{\scshape i\kern-0.25em b}\kern-0.8em\TeX}}}

\usepackage{amssymb}
\usepackage{sidecap}
\usepackage{booktabs}
\usepackage{amsmath} 
\usepackage{xcolor}
\usepackage{enumitem}
\usepackage{multirow} 
\usepackage{algorithm}
\usepackage{algpseudocode}
\usepackage{xcolor}
\usepackage{url}
\usepackage{subfloat}
\usepackage{subfig}
\usepackage{xurl}
\usepackage{pifont}

\usepackage{tikz}

\usepackage[subtle,tracking=normal, title=normal]{savetrees}
\usepackage{booktabs}  
\usepackage{makecell}  
\usepackage[table]{xcolor} 
         
\setcopyright{none}                
\makeatletter
\def\@copyrightspace{\relax}       
\makeatother

\acmConference[DAC '26]{The 63rd Design Automation Conference}{June 2026}{Long Beach, CA, USA}
\acmYear{2026}
\acmDOI{}
\acmISBN{}

\begin{document}
\settopmatter{printfolios=true}
\settopmatter{printacmref=false} 

\title{SecureRouter: Encrypted Routing for Efficient Secure Inference}

\author{Yukuan Zhang} 
\affiliation{%
  \institution{University of Central Florida}
  \city{Orlando}
  \state{Florida}
  \country{USA}
}
\email{yukuan.zhang@ucf.edu}

\author{Mengxin Zheng} 
\affiliation{%
  \institution{University of Central Florida}
  \city{Orlando}
  \state{Florida}
  \country{USA}
}
\email{mengxin.zheng@ucf.edu}

\author{Qian Lou}
\affiliation{%
  \institution{University of Central Florida}
  \city{Orlando}
  \state{Florida}
  \country{USA}
}
\email{qian.lou@ucf.edu}

\begin{abstract}
Cryptographically secure neural network inference typically relies on secure computing techniques such as Secure Multi-Party Computation (MPC), enabling cloud servers to process client inputs without decrypting them. Although prior privacy-preserving inference systems co-design network optimizations with MPC, they remain slow and costly, limiting real-world deployment. A major bottleneck is their use of a single, fixed transformer model for all encrypted inputs, ignoring that different inputs require different model sizes to balance efficiency and accuracy. We present \textit{SecureRouter}, an end-to-end encrypted routing and inference framework that accelerates secure transformer inference through input-adaptive model selection under encryption. SecureRouter establishes a unified encrypted pipeline that integrates a secure router with an MPC-optimized model pool, enabling coordinated routing, inference, and protocol execution while preserving full data and model confidentiality. The framework includes training-phase and inference-phase components: an MPC-cost-aware secure router that predicts per-model utility and cost from encrypted features, and an MPC-optimized model pool whose architectures and quantization schemes are co-trained to minimize MPC communication and computation overhead. Compared to prior work, SecureRouter achieves a latency reduction by 1.95$\times$ with negligible accuracy loss, offering a practical path toward scalable and efficient secure AI inference. Our open-source implementation is available at : \url{https://github.com/UCF-ML-Research/SecureRouter}

\end{abstract}

\maketitle

\section{Introduction}

State-of-the-art Transformer models such as XLNet~\cite{yang2019xlnet} and ALBERT~\cite{lan2019albert} have become indispensable in modern NLP applications. However, deploying these models in sensitive domains---including medicine and finance---raises serious privacy concerns. To address these risks, Secure Multi-Party Computation (MPC) has emerged as a foundational technique for Privacy-Preserving Machine Learning (PPML)~\cite{gilad2016cryptonets, mohassel2017secureml, liu2017minionn, knott2021crypten}, enabling cloud servers to process client inputs without ever decrypting them. In this setting, clients secret-share their data across two or more non-colluding servers, which collaborate to execute the inference protocol directly over encrypted shares. This design ensures that neither the raw inputs nor the model parameters are ever exposed in plaintext.

\begin{figure}[t]
    \centering
    \includegraphics[width=\columnwidth]{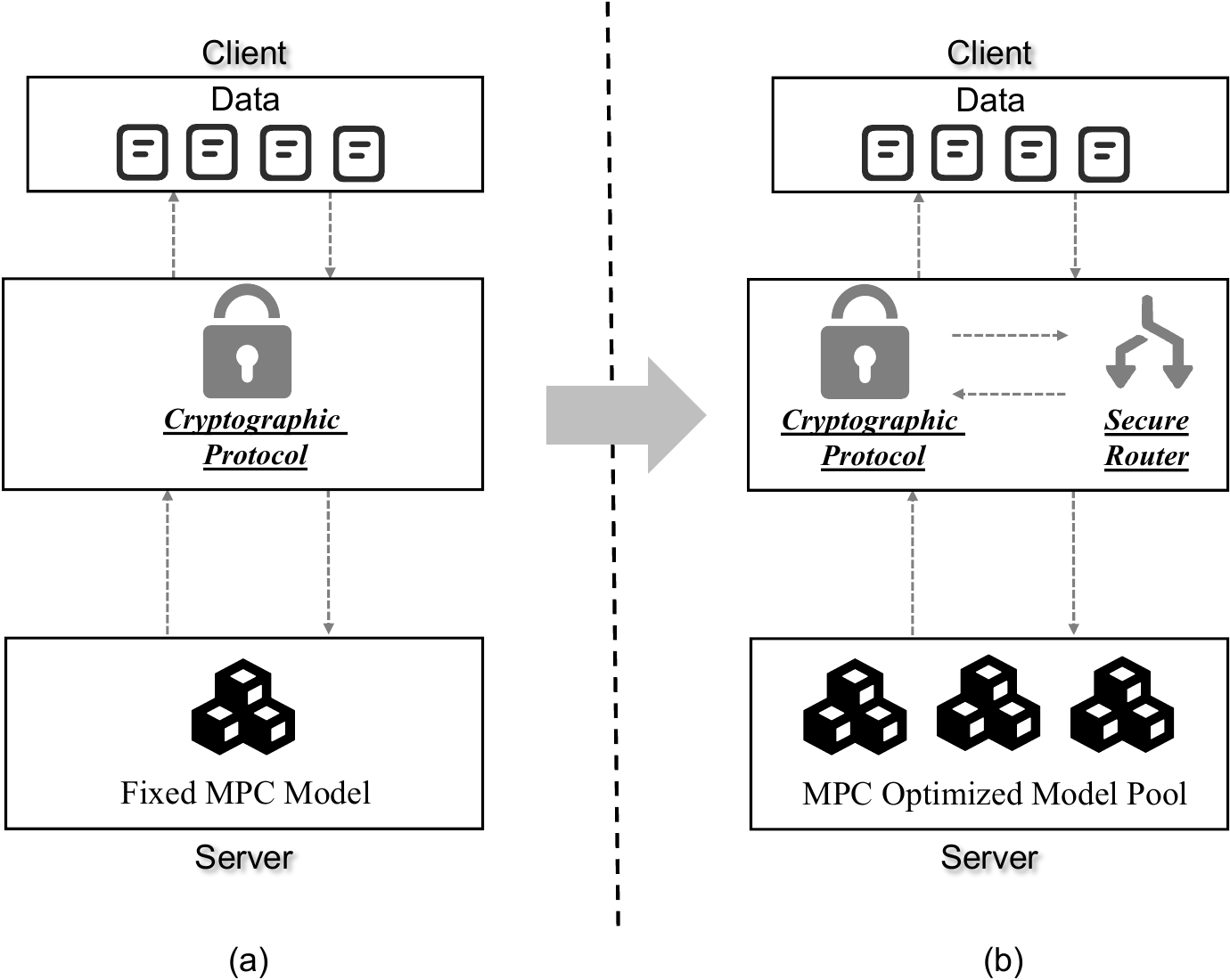}
    \caption{Comparison of (a) a current MPC framework utilizing a single, fixed model, with (b) our proposed framework, which introduces a Secure Router to dynamically select an appropriate model from an MPC Model Pool.}
    \label{fig:introduction}
\end{figure}

Despite these strong confidentiality guarantees, the practical adoption of MPC-based inference remains limited due to its prohibitive computational cost. The dominant bottleneck arises from non-linear operations such as GeLU and Softmax, which are trivial in plaintext but account for over 77\% of secure inference time~\cite{luo2024secformer}. As a consequence, a standard \texttt{BERT$_{\text{BASE}}$} model that runs in under one second in plaintext can exceed 60 seconds under MPC~\cite{li2023mpcformer}. Recent frameworks including MPCFormer~\cite{li2023mpcformer} and SecFormer~\cite{luo2024secformer} alleviate part of this overhead by optimizing or approximating non-linear components. However, these systems still follow a \emph{single-model} design, as shown in Figure~\ref{fig:introduction}(a): every encrypted input is processed by the same fixed Transformer, regardless of its difficulty or computational demand. This paradigm forces all queries---easy and hard alike---through the same expensive computation and fails to resolve the core trade-off between the accuracy of larger models and the latency benefits of smaller ones.

In plaintext inference, this inefficiency is commonly resolved using \emph{input-adaptive inference}, or \emph{model routing}~\cite{jitkrittum2025universal, shirkavand2025cost}. Instead of relying on a single model, plaintext systems maintain a pool of heterogeneous LLMs and use a lightweight router to select, per input, the smallest model capable of maintaining accuracy. However, these methods are fundamentally unsuitable for MPC: they assume full plaintext visibility, and their cost metrics rely on FLOPs or API pricing, whereas MPC cost is dominated by communication rounds, cryptographic primitives, and non-linear operations—not parameter count alone.

A key challenge in extending routing to the encrypted setting is that both training and inference must operate under MPC constraints. During training, the framework must construct MPC-optimized model pools whose architectures and quantization schemes are optimized for secure computation, rather than plaintext FLOPs. The router itself must learn to predict model utility and cost from statistics derived from encrypted data, despite never observing raw inputs. At the same time, the system must define an MPC-relevant cost model---capturing communication overhead, multiplication counts, and non-linear activation costs---that meaningfully differs from plaintext cost. During inference, these components must be tightly integrated: the router must operate on secret-shared embeddings, its decisions must remain encrypted, and the selected model must be retrieved obliviously and executed within the MPC protocol. Designing a pipeline that combines these elements efficiently, without revealing the input or the routing choice, forms the central technical difficulty of encrypted routing.

To bridge this gap, we present \textbf{SecureRouter}, an end-to-end encrypted routing and inference framework that accelerates secure Transformer inference through input-adaptive model selection under encryption. As illustrated in Figure~\ref{fig:introduction}(b), SecureRouter introduces a \emph{Secure Router} that adaptively selects the most efficient model from a secret-shared MPC Model Pool without revealing either the input or the routing decision. SecureRouter establishes a unified encrypted pipeline that integrates (1) an MPC-cost-aware router capable of predicting per-model utility and cost directly from encrypted features, and (2) an MPC-optimized model pool whose architectures and quantization schemes are co-trained to minimize secure communication and computation overhead.

\noindent\textbf{Contributions.} Our contributions are listed as follows:
\begin{itemize}
    \item \textbf{Unified Encrypted Routing Framework.} We design the first end-to-end MPC-based routing pipeline that jointly orchestrates encrypted routing, oblivious model retrieval, and secure Transformer inference while preserving full input, decision, and model confidentiality.
    \item \textbf{MPC-Cost-Aware Router and Model Pool Co-Design.} We introduce an MPC-cost-aware router trained to predict model utility and execution cost from encrypted statistics, and a co-trained model pool optimized to reduce MPC communication and non-linear computation overhead.
    \item \textbf{Practical Efficiency Gains.} SecureRouter achieves up to \textbf{1.95$\times$ lower inference latency} with negligible accuracy loss across GLUE tasks compared to fixed-model MPC baselines, providing a practical path toward scalable and efficient privacy-preserving Transformer inference.
\end{itemize}

\section{Background and Related Work}
\subsection{Threat Model}
Consistent with established privacy-preserving inference frameworks, we adopt the standard 
semi-honest (or honest-but-curious) threat model \cite{goldreich1987how, lou2019she, zhang2025cipherprune, zhang2024heprune, lou2021safenet}. In this setting, all participating parties—including the client and the computing servers—strictly adhere to the specified protocol instructions and do not tamper with the data or computation. 
However, they are considered adversarial in that they may inspect execution transcripts and intermediate memory states in an attempt to infer sensitive information about the private inputs or model parameters.

\subsection{Secure Multi-party Computation for Neural Networks}
Secure Multi-party Computation provides a cryptographic framework for multiple parties to collaboratively compute a function over their private data without revealing the inputs to one another. The robust privacy guarantees protect both the data and model weights. A significant body of work has explored the implementation of various neural networks, including Transformer models, within secure computation frameworks \cite{mohassel2017secureml, mohassel2018aby3, riazi2018chameleon, wagh2019securen, mishra2020delphi, knott2021crypten}.  Rather than developing a new MPC system, this paper introduces an end-to-end MPC router designed to accelerate Transformer inference, with the goal of portability across existing MPC frameworks.

To enforce this privacy, the existing frameworks relies on secret sharing schemes \cite{goldreich2019play, damgard2012multiparty}. We specifically utilize additive secret sharing over a ring $\mathbb{Z}_L$. In this scheme, a secret value $x$ is decomposed into two random shares, $\langle x \rangle_1$ and $\langle x \rangle_2$, which are distributed to the user and the model provider, respectively. This decomposition satisfies $x = \langle x \rangle_1 + \langle x \rangle_2 \pmod L$. This mechanism ensures information-theoretic security, as possessing a single share reveals no information regarding the underlying secret $x$, while the combination of shares allows for the reconstruction of the original value.

While linear operations (such as addition) can be computed locally by summing individual shares, non-linear operations like multiplication require cryptographic protocols involving communication. A standard approach for secure multiplication utilizes Beaver triples \cite{beaver1991efficient}. Given secret-shared inputs $x$ and $y$, and a pre-generated secret-shared triple $(a, b, c)$ where $c = ab$, the parties compute the masked differences $\epsilon = x - a$ and $\delta = y - b$ locally. These masked values are exchanged to reconstruct the plaintext $\epsilon$ and $\delta$. The product $z = xy$ is then computed as a linear combination of the public masked values and the private shares: $z = c + \epsilon b + \delta a + \epsilon \delta$.

Beyond arithmetic computation, conditional execution on private data 
requires Oblivious Transfer (OT) \cite{rabin1981how, even1985randomized}. 
OT is a fundamental cryptographic primitive that allows a receiver to 
select and retrieve a specific element from a dataset of sender without 
revealing the selection index to the sender, while ensuring the receiver 
gains no information about the unselected elements. Once our 
secure router determines the optimal expert index in a secret-shared 
format, we utilize OT-based protocols to selectively retrieve the 
corresponding model parameters from the encrypted model pool.

\subsection{The Transformer Pre-training and Fine-tuning Paradigm}
The Transformer architecture has become a cornerstone of modern machine learning, demonstrating state-of-the-art performance across diverse domains. Initially revolutionizing Natural Language Processing \cite{yang2019xlnet, lan2019albert, raffel2020t5, clark2020electra, lou2022dictformer, hsu2022language}, its principles have since been successfully adapted for computer vision \cite{dosovitskiy2020vit, liu2021swin, radford2021clip} and other modalities \cite{sharir2021videoar, lou2022lite}.
A dominant paradigm for the application of these models is a \textbf{two-stage} strategy: (1) pre-training on a massive, general-purpose dataset to learn broad data representations, followed by (2) fine-tuning on a smaller, task-specific downstream dataset. This pre-training and fine-tuning methodology has proven highly effective and is widely adopted \cite{radford2018gpt, liu2019roberta, turc2019exploring, wolf2019transformers}.

\subsection{Routing-based Inference}
The Mixture of Experts (MoE) architecture is a form of conditional computation 
designed to increase model capacity without a proportional rise in computational 
cost \cite{shazeer2017outrageously, jiang2024mixtral, fedus2022switch}. A 
trainable gating network or router~\cite{xue2026r2} learns to dynamically select a sparse subset 
of these experts (e.g., the top 1 or 2) for input. While classical MoE 
implementations typically involve equally-sized internal sub-models, modern 
approaches employ sparse activation strategies to rigorously minimize 
computational overhead \cite{shirkavand2025cost}. This paradigm has recently 
been extended to coarse-grained MoEs, where each expert constitutes a full, 
standalone Large Language Model (LLM) rather than an internal sub-layer, thereby 
enabling the dynamic routing of inputs across a heterogeneous pool of complete 
models \cite{shirkavand2025cost, jitkrittum2025universal}. This strategy avoids over-paying for simple tasks, ensuring that 
expensive, high-cost models are used sparingly, only on the 
(relatively) few hard inputs \cite{jitkrittum2025universal}. 
This concept can be viewed as a coarse-grained MoE (Mixture of Experts), 
where each expert is a full LLM. 
Pioneering MoE architectures like the Switch Transformer 
\cite{fedus2022switch} and Mixtral \cite{jiang2024mixtral} 
leverage similar sparse routing mechanisms internally to reduce 
computation. 

\subsection{Gumbel-Softmax}
A fundamental challenge in training routing mechanisms, such as those in MoE models, is the need to make discrete, non-differentiable choices (e.g., select expert 1 or select expert 3). Gradient-based optimization via backpropagation, however, requires a continuous and differentiable computation path. The Gumbel-Softmax trick, also introduced concurrently as the Concrete distribution \cite{jang2017gumbel, maddison2017concrete}, provides a solution. It is a reparameterization technique that creates a continuous and differentiable relaxation of a discrete categorical distribution.

\section{SecureRouter}
In this section, we present the SecureRouter framework, an end-to-end system for privacy-preserving inference. The core of this framework, illustrated in Figure \ref{fig:secure_router}, is to employ a cost-aware router in MPC environment that dynamically selects the optimal model from a MPC optimized model pool based on the specific query.

\subsection{System Overview}
We will begin with a high-level \textbf{Framework} which shows the workflow of our SecureRouter. Following this, we will examine the specific mechanics of the secure \textbf{Inference Protocol} used during the online phase and the \textbf{Generator (training)} process used to build the router which balances accuracy and cost after training in the offline phase.

\subsection{SecureRouter Framework}
\begin{figure}[htbp]
    \centering
    \includegraphics[width=\columnwidth]{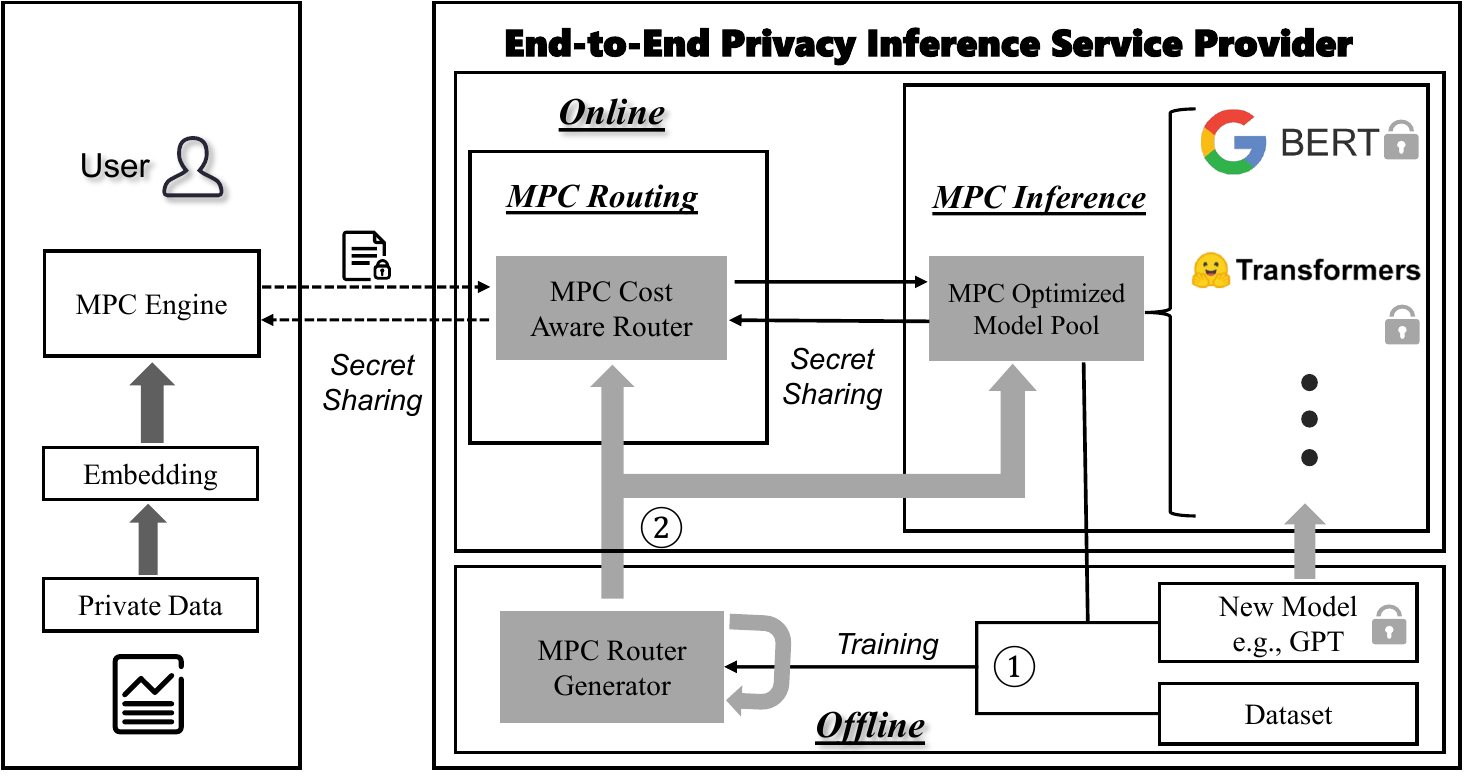}
    \caption{An illustration of our proposed secure router framework, divided into an offline training phase and an online inference phase. The diagram simplifies the architecture to focus on the User and the End-to-End Privacy Inference Service Provider.}
    \label{fig:secure_router}
\end{figure}

As illustrated in Figure \ref{fig:secure_router}, SecureRouter establishes a unified encrypted pipeline operating in two distinct phases. During the \textbf{offline phase}, the system focuses on optimizing the components for the specific constraints of secure computing. In step \textcircled{1}, the \textbf{MPC-cost-aware secure router} is trained to predict per-model utility and execution cost directly from features. Parallel to this, in step \textcircled{2}, the \textbf{MPC-optimized model pool} is constructed. Here, diverse transformer architectures are co-trained and quantized to explicitly minimize MPC communication and computation overhead, before being converted into secret-shared formats for secure deployment.

Transitioning to the \textbf{online phase}, the framework executes input-adaptive secure inference. When the system receives secret-shared input embeddings from the user, the router dynamically selects the optimal model index from the pool based on the input's characteristics and the learned cost-utility policy. Subsequently, the selected MPC-optimized model executes the inference protocol. This coordinated pipeline ensures that the user’s private input, the routing decision, and the model parameters remain cryptographically confidential throughout the computation, with the final encrypted results returned via the MPC Engine.

\subsection{SecureRouter Inference Protocol}
\begin{figure}[htbp]
    \centering
    \includegraphics[width=\columnwidth]{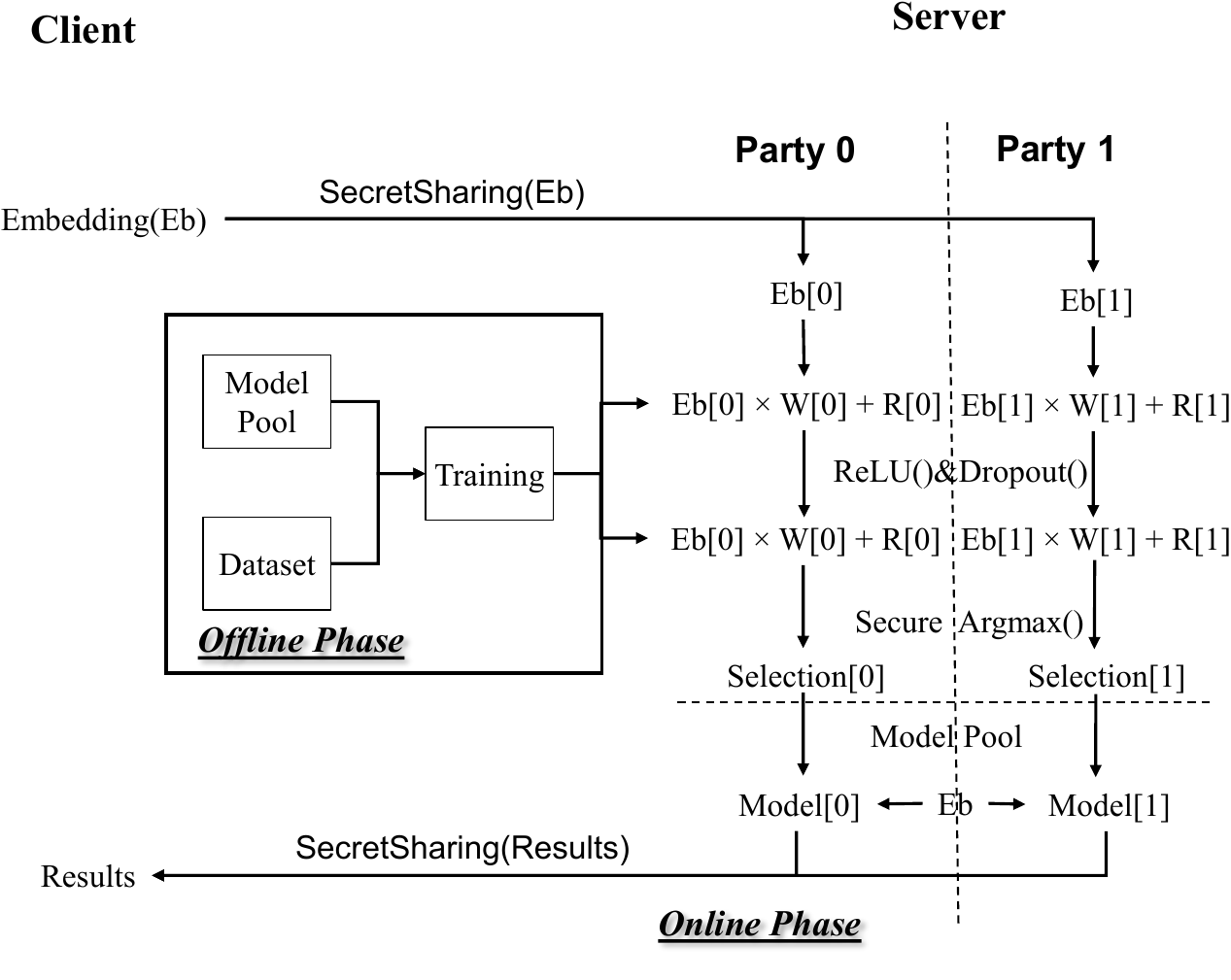}
    \caption{The online inference protocol, illustrating the 2-Party Computation (2PC) flow between the \texttt{Client} and the \texttt{Server} (\texttt{Party 0} and \texttt{Party 1}).}
    \label{fig:Protocol}
\end{figure}

The online inference protocol, detailed in Figure \ref{fig:Protocol}, commences when the client processes their private data to generate an embedding $\mathbf{e}$. This embedding is then transformed into secret shares $[\mathbf{e}]_0$ and $[\mathbf{e}]_1$. These shares are distributed to two non-colluding server parties, Party 0 and Party 1, respectively.

Upon receiving the shares, the server parties collaboratively execute the privacy-preserving routing mechanism using a pre-trained router policy from the offline phase. Specifically, Party 0 computes $[\mathbf{e}]_0 \times [\mathbf{W}]_0 + [\mathbf{R}]_0$ while Party 1 computes $[\mathbf{e}]_1 \times [\mathbf{W}]_1 + [\mathbf{R}]_1$, where $[\mathbf{W}]$ and $[\mathbf{R}]$ are the secret-shared weights and biases of the router policy learned during the offline training phase. These linear transformations are followed by secure non-linear activations (ReLU and Dropout), forming a secret-shared two-layer neural network that evaluates routing decisions. The outputs from both parties are then fed into a secure argmax protocol, which cryptographically computes the index of the optimal model.

The selection result, maintained in secret-shared format ($[Selection]_0$ and $[Selection]_1$), is used to retrieve the corresponding model parameters from the MPC optimized model pool via oblivious transfer. The system then executes the main MPC inference using the original embedding shares ($[\mathbf{e}]_0$ and $[\mathbf{e}]_1$) and the privately retrieved model parameters ($[\mathbf{Model}]_0$ and $[\mathbf{Model}]_1$). The final inference results remain in secret-shared state ($[\mathbf{y}]_0$ and $[\mathbf{y}]_1$) and are sent back to the client. Only the client, upon recombining the shares, can reconstruct the final plaintext result.

\subsection{SecureRouter Generator}
\begin{figure}[htbp]
    \centering
    \includegraphics[width=\columnwidth]{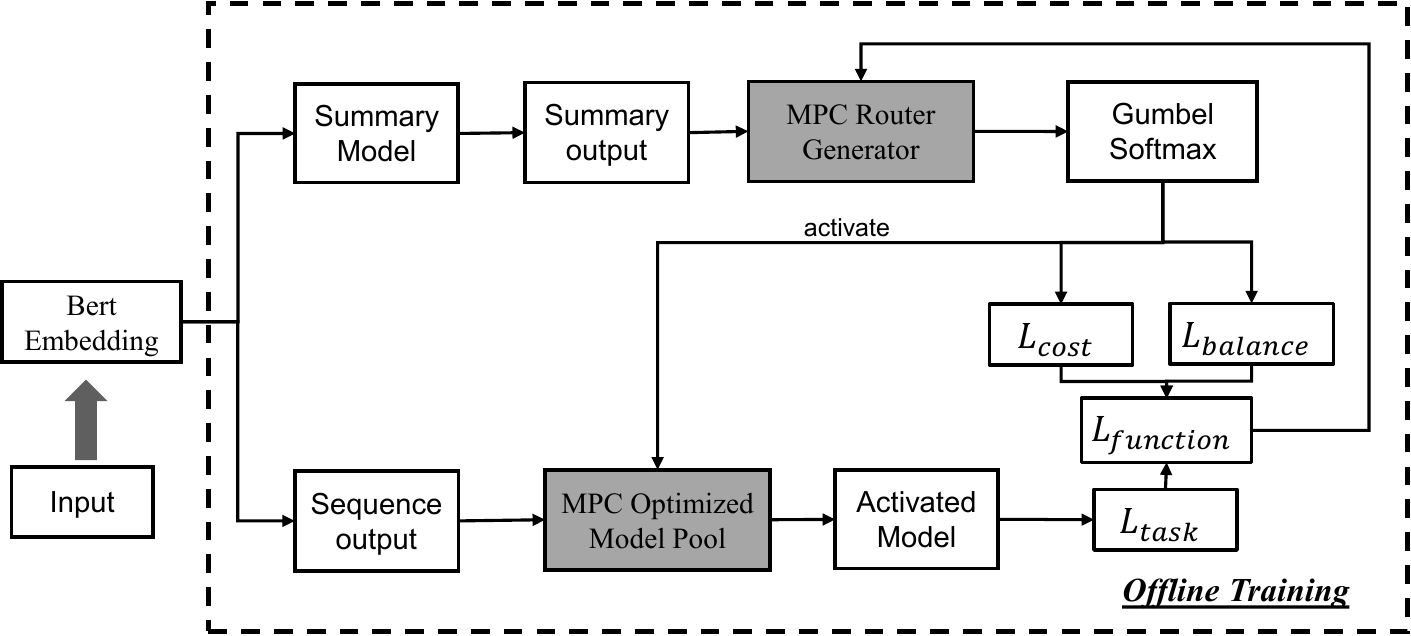}
    \caption{The offline training architecture for the Router Generator. The BERT embedding trains the MPC-cost-aware aware Router via a Summary Model and serves as input to the MPC optimized model pool. The router's Gumbel Softmax selection computes routing losses ($L_{cost}$, $L_{balance}$), activates an expert model, and produces task loss ($L_{task}$) for joint router-expert training.}
    \label{fig:router_generator}
\end{figure}

The offline training process for the Router Generator is illustrated in Figure \ref{fig:router_generator}. This phase is critical for the joint optimization of the MPC-cost-aware Router (e.g., BERT Tiny + a lightweight MLP) and the MPC-optimized Model Pool. As highlighted in our contributions, this process co-trains the router to make cost-aware decisions while simultaneously tuning the expert architectures and their quantization parameters to minimize MPC overhead.

The training data flow commences when an input is processed to generate a BERT embedding $\mathbf{e}$. This embedding is subsequently utilized along two parallel computational branches. 

In the expert path, the complete embedding sequence is forwarded to the MPC optimized model pool, where it becomes accessible for processing by any of the expert models (e.g., different BERT variants). Concurrently, in the router path, the embedding is first passed through a summary model(e.g. BERT Tiny) to produce a condensed representation. This summarized output serves as the input to the MPC Router Generator, which generates selection logits. To enable end-to-end differentiable training, these logits are processed through a Gumbel Softmax function, providing a continuous relaxation of the discrete expert selection process and facilitating gradient-based optimization of the router's decision-making mechanism via backpropagation.

The Gumbel Softmax output is used to calculate a composite loss function ($L_{function}$) and to produce the final task prediction. The training is guided by three distinct loss components, as detailed in the provided images.

\subsubsection{The Main Task Loss ($L_{task}$)}

The primary objective function ensures that the integrated system—comprising both the router and the expert models—achieves accurate classification performance. The final prediction $Y_{pred}$ is computed as a weighted aggregation of outputs from all experts $E_i$ in the MPC optimized model pool, where the weights $g_i$ correspond to the probabilities generated by the router's Gumbel Softmax operation:
\begin{equation}
Y_{pred} = \sum_{i=1}^{k} g_i \cdot E_i(x)
\label{eq:weighted_prediction}
\end{equation}
where $k$ denotes the total number of expert models, $E_i(x)$ represents the output of the $i$-th expert model, and $g_i$ is the routing probability for expert $i$.

The task loss is computed using the cross-entropy loss between the predicted distribution and the ground truth labels:
\begin{equation}
L_{task} = \text{CrossEntropyLoss}(Y_{pred}, Y_{true})
\label{eq:task_loss}
\end{equation}

\subsubsection{The Load Balancing Loss ($L_{balance}$)}

The load balancing objective is designed to promote uniform utilization of all expert models by the MPC Router across a training batch, thereby preventing mode collapse where the router converges to selecting only a subset of available experts. This loss is computed using the squared coefficient of variation of the expert utilization distribution.

For each expert $i \in \{1, \ldots, k\}$, the aggregate load $L_i$ over a batch $B$ is defined as the sum of routing probabilities assigned to that expert across all samples:
\begin{equation}
L_i = \sum_{x \in B} g_i(x)
\label{eq:expert_load}
\end{equation}
where $g_i(x)$ denotes the routing probability assigned to expert $i$ for input $x$.

The load balancing loss is then formulated as the squared coefficient of variation of the expert loads, quantifying the relative dispersion of utilization across experts:
\begin{equation}
L_{balance} = \frac{\text{Var}(L_1, \ldots, L_k)}{\left[\text{Mean}(L_1, \ldots, L_k)\right]^2}
\label{eq:balance_loss}
\end{equation}

This formulation penalizes uneven expert utilization, encouraging the router to maintain a balanced distribution of samples across the MPC optimized model pool during training.

\subsubsection{The Cost-Aware Loss ($L_{cost}$)}

The cost-aware objective function incentivizes the MPC Router to favor computationally efficient expert models during selection. To reflect the computational constraints of secure multi-party computation, we define the cost metric $c_i$ for each expert $i$ as the empirical inference time measured within the MPC environment. For instance, smaller models incur lower communication and computation costs in MPC execution (e.g., $c_{\text{BERT-small}} = 1.0$, $c_{\text{BERT-base}} = 2.5$).

For a given input $x$, the expected computational cost is formulated as the weighted sum of individual expert costs, where the weights correspond to the routing probabilities:
\begin{equation}
\text{ExpectedCost}(x) = \sum_{i=1}^{k} g_i(x) \cdot c_i
\label{eq:expected_cost}
\end{equation}
where $g_i(x)$ denotes the routing probability assigned to expert $i$ for input $x$, and $c_i$ represents the MPC inference time cost of expert $i$.

The cost-aware loss is then computed as the mean expected cost over the entire training batch $B$:
\begin{equation}
L_{cost} = \frac{1}{|B|} \sum_{x \in B} \text{ExpectedCost}(x)
\label{eq:cost_loss}
\end{equation}

As illustrated in Figure \ref{fig:router_generator}, the load balancing loss $L_{balance}$ and the cost-aware loss $L_{cost}$ are aggregated into a unified routing objective $L_{function}$, which, in conjunction with the task loss $L_{task}$, governs the joint optimization of both the router and expert models throughout the offline training phase.

\section{Experiments}

This section showcases the effectiveness of SecureRouter through experiments. We begin with the \textbf{experiment setup} and then report the performance assessment results in \textbf{performance comparison}.
\subsection{Experimental Setup}

\textbf{Implementation.} \ \ Our system, SecureRouter, was implemented using the Crypten framework, a semi-honest, secret-sharing-based platform for privacy-preserving machine learning.
The hardware configurations for our experiments are offline and online phases. The \textbf{offline training} and router generation were performed on a single server equipped with an INTEL(R) XEON(R) GOLD 6526Y CPU, 32GB of system memory, and one NVIDIA H100 GPU. The \textbf{online inference} protocol was evaluated in a simulated two-party computation (2PC) environment, which comprised two local servers. Each of these servers was equipped with an NVIDIA 3090 GPU, and the two parties were interconnected via a 10 Gbps network.\\
\textbf{Models.} \ \ The experimental models in MPC optimized model pool are standard BERT architectures sourced from the HuggingFace Transformers library, selected to represent a range of computational scales. The most compact model, BERT-tiny, features 2 Transformer encoder layers, a hidden size of 128, 2 attention heads, and approximately 4.4 million parameters. The foundational BERT-Base version comprises 12 layers, a hidden size of 768, 12 attention heads, and 110 million parameters. Finally, BERT-Large, an expanded iteration, is configured with 24 layers, a hidden size of 1024, 16 attention heads, and approximately 340 million parameters, enabling it to capture more intricate language patterns.\\
\textbf{Datasets.} \ \ To ensure a comprehensive and reliable evaluation, we conduct experiments across a diverse suite of benchmarks drawn from the General Language Understanding Evaluation (GLUE) dataset \cite{wang2019glue}. This collection includes the MNLI, QQP, SST-2, RTE, MRPC, CoLA, STS-B, and QNLI tasks, which intentionally span a variety of evaluation metrics. Specifically, performance is measured using accuracy (for MNLI, RTE, SST-2, and QNLI), the F1 score (for MRPC and QQP), the Matthews correlation coefficient (for CoLA), and the average of Pearson and Spearman correlations (for STS-B).\\
\textbf{Baselines for Comparison.} \ \ 
We establish two primary sets of baselines for our experimental evaluation. The first baseline consists of a standard Hugging Face Transformer model \cite{frantar2021m, matsubara2023torchdistill, he2021distillersystematicstudymodel} fine-tuned on the target dataset, with its secure inference performance evaluated using the Crypten framework. The second set of baselines follows the setting in SecFormer \cite{luo2024secformer}, for which we report the individual, fine-tuned performance of each redesigned model that is included in our MPC optimized model pool.

\subsection{Performance Comparison}

We present our primary results in Table \ref{tab:results}. Comparing our MPC-cost-aware Router + MPC optimized model pool with the BERT-Large fine-tuned baseline, our framework exhibits a significant performance improvement.

\begin{table}[htbp]
  \centering

  \resizebox{\columnwidth}{!}{%
    \begin{tabular}{lccccccccc}
      \toprule
      \textbf{Method} & 
      \textbf{\makecell{MNLI \\ 393k}} & 
      \textbf{\makecell{QQP \\ 363k}} & 
      \textbf{\makecell{QNLI \\ 108k}} & 
      \textbf{\makecell{SST-2 \\ 67k}} & 
      \textbf{\makecell{CoLA \\ 8.5k}} & 
      \textbf{\makecell{STS-B \\ 5.7k}} & 
      \textbf{\makecell{MRPC \\ 3.5k}} & 
      \textbf{\makecell{RTE \\ 2.5k}} & 
      \textbf{\makecell{Average \\ Speed-up}} \\
      \midrule

      \makecell[l]{BERT-Large \\ fine-tuned} & 
      86.64 & \textbf{88.08} & \textbf{92.24} & \textbf{93.46} & \textbf{63.35} & \textbf{90.36} & 91.62 & \textbf{75.45} & 1x \\
      \midrule
      
      \makecell[l]{SecureRouter \\ + MPC optimized model pool} & 
      \makecell{\textbf{86.73} \\ (1.21x)} & 
      \makecell{88.05 \\ (1.95x)} & 
      \makecell{92.00 \\ (1.20x)} & 
      \makecell{93.10 \\ (1.73x)} & 
      \makecell{59.10 \\ (1.22x)} & 
      \makecell{89.02 \\ (1.30x)} & 
      \makecell{\textbf{91.78} \\ (1.16x)} & 
      \makecell{75.09 \\ (1.49x)} & 
      1.53x \\
      
      \bottomrule
    \end{tabular}%
  } 
    \caption{Performance comparison of BERT Large and SecureRouter. Bolded numbers indicate best results;Values in parentheses represent the inference speed-up of our MLP Router + MPC optimized model pool method relative to the BERT-Large fine-tuned baseline for each task.}
    \label{tab:results}
\end{table}
For instance, on MNLI, our model (86.73) slightly outperforms the baseline (86.64), and on QQP, it achieves a 1.95x speed-up with a negligible performance trade-off (88.05 vs. 88.08). The most significant speed-ups are observed on QQP (1.95x) and SST-2 (1.73x), validating our approach's ability to dynamically allocate resources and accelerate secure inference with minimal impact on model quality.

Furthermore, the results highlight a distinct relationship between task characteristics and router effectiveness. Tasks centered on semantic similarity and inference, such as MRPC and QQP, benefit significantly from our architecture, yielding either unexpected performance gains (as seen in MRPC, +0.16) or massive speed-ups (1.95x on QQP). Conversely, the performance dip observed on CoLA (59.10 vs. 63.35) suggests that tasks requiring strict syntactic and grammatical precision may be more sensitive to the reduced capacity of selected models in the pool. Despite this specific trade-off, the method’s ability to increase accuracy on the massive MNLI dataset (+0.09) while providing a 1.21x speed-up confirms that the SecureRouter effectively identifies 'easy' samples in large-scale semantic workloads, optimizing the cost-accuracy pareto frontier.

\begin{table}[htbp]
  \centering

  \resizebox{\columnwidth}{!}{%
    \begin{tabular}{lccccccccc}
      \toprule
      \textbf{Method} & 
      \textbf{\makecell{QNLI \\ 108k}} & 
      \textbf{\makecell{CoLA \\ 8.5k}} & 
      \textbf{\makecell{STS-B \\ 5.7k}} & 
      \textbf{\makecell{MRPC \\ 3.5k}} & 
      \textbf{\makecell{RTE \\ 2.5k}} & 
      \textbf{\makecell{Average \\ Speed-up}} \\
      \midrule

      \makecell[l]{SecFormer \\ BERT-Large} & 
      37.75s & 37.75s & 37.75s & 37.75s & 37.75s & 1x \\
      \midrule
      
      \makecell[l]{SecureRouter \\ + MPC optimized model pool} & 
      \makecell{19.6s \\ (1.92x)} & 
       \makecell{18.50s \\ (2.04x)}  & 
      \makecell{20.62s \\ (1.83x)} & 
      \makecell{19.26s \\ (1.96x)} & 
      \makecell{17.23s \\ (2.19x)} & 
      1.95x \\
      
      \bottomrule
    \end{tabular}%
  } 
    \caption{\textbf{Inference Latency and Speed-up Comparison on GLUE Benchmarks.} Following the experimental setting of SecFormer \cite{luo2024secformer}, we measure the end-to-end inference time (in seconds) for a single sample.}
    \label{tab:results}
\end{table}
SecureRoute consistently outperforms the state-of-the-art SecFormer framework across all evaluated tasks, reducing the average inference latency by nearly 50\% while operating within the same secure experimental constraints.
Extensive profiling reveals that the extent of the speed-up is inversely correlated with the inherent difficulty of the dataset, confirming the router's ability to successfully exploit sample redundancy. It not only achieves a high efficiency gain of 2.19$\times$ on simpler tasks like RTE by utilizing smaller experts but also maintains strict accuracy on semantically complex tasks like STS-B, necessitating a more conservative speed-up of 1.83$\times$. SecureRoute demonstrates a sophisticated capacity for adaptive resource allocation, offering a dynamic solution that promises to minimize MPC overhead on easy samples while automatically scaling to meet the rigorous demands of challenging inputs.

\subsection{Time efficiency}

 Given the significant computational overhead of conducting end-to-end MPC inference for every sample in the GLUE benchmark, we report the projected speed-up based on router profiling. To calculate this, we first measured the unitary inference latency for each expert model (Tiny, Base, Large) within our MPC environment. We then ran the SecureRoute router on the test set to obtain the distribution of expert selections for each task. The average speed-up is calculated as the ratio of the baseline cost (running BERT-Large for all $N$ samples) to the weighted sum of the costs of the selected experts, plus the overhead of the router itself:

$$\text{Speed-up} = \frac{N \times C_{\text{Large}}}{\sum_{i=1}^{N} (C_{\text{selected}^{(i)}} + C_{\text{router}})}$$
where $C_{\text{Large}}$ is the MPC latency of the baseline model, and $C_{\text{selected}^{(i)}}$ is the latency of the model chosen by the router for the $i$-th sample.
For example, in the result of RTE dataset, in Table \ref{tab:rte}, We can find that our SecureRouter  is significantly more efficient than the BERT-Large fine-tuned baseline. Specifically, our approach achieves an average running time of 133.92s in each result in the evaluation set, which is 1.49x faster than the baseline's 199.78s. This significant speedup is achieved with only a minimal 0.36\% drop in accuracy (75.09\% vs. 75.45\%).

\begin{table}[h]

\centering
\resizebox{\columnwidth}{!}{%
\begin{tabular}{lccc}
\toprule
Method & Accuracy (\%) & \begin{tabular}[c]{@{}c@{}}Expert Distribution\\ {[}Tiny, base, Large{]}\end{tabular} & \begin{tabular}[c]{@{}c@{}}Running\\ Time(s)\end{tabular} \\ 
\midrule
\begin{tabular}[c]{@{}l@{}}BERT Large\\ fine-tuned\end{tabular} & 75.45 & {[}0,0,100{]} & 199.78 \\ 
\begin{tabular}[c]{@{}l@{}}SecureRouter\\ + MPC optimized model pool\end{tabular} & 75.09 & {[}16.9, 28.3, 54.8{]} & 133.92 \\ 
\bottomrule
\end{tabular}%
}
\caption{Performance comparison on RTE dataset. Running time is the average of time for the evaluation with fine-tuned BERT Large.}
\label{tab:rte}
\end{table}

These advantages stem from our router-based MPC optimized model pool. Instead of processing every sample with the full BERT-Large model (which has an expert distribution of [0, 0, 100]), our MPC-cost-aware Router dynamically distributes the workload. As shown in the expert distribution \textbf{[16.9, 28.3, 54.8]}, the model is able to route a large portion of samples to the more efficient Tiny and Base experts, with only 54.8\% of samples requiring the full BERT-Large model. This dynamic routing significantly reduces the average inference cost.
\begin{table}[htbp]
    \centering 
      \resizebox{\columnwidth}{!}{%
    \begin{tabular}{lcccc}
        \toprule 
        \textbf{Model} & \textbf{SecureRouter} & \textbf{BERT\_tiny} & \textbf{BERT\_base} & \textbf{BERT\_large} \\
        \midrule 
        
        Running Time(s) & 4.17 & 4.11 & 69.71 & 199.78 \\
        Communication\\ Volume(GB) & 1.86 & 1.86 & 58.50 & 156.31 \\
        Memory \\ Usage(MB) & 39.12 & 38.00 & 1054.00 & 3114.00\\
        \bottomrule 
    \end{tabular}
    }
    \caption{Comparison of baseline Model Running Time(s), Communication Volume(GB) and Memory Usage(MB) with SecureRouter in 2-PC environment}
    \label{tab:running_times}
\end{table}
A detailed time breakdown of the individual components is shown in Table \ref{tab:running_times}. This table lists the running times for the SecureRouter (4.17s) with 1.86 GB communication volume as well as the inference costs for each expert in the pool: BERT tiny (4.11s) with 1.86 GB communication volume, BERT base (69.71s) with 58.50 GB communication volume, and BERT large (199.78s) with 156.31GB communication volume.
Critically, the memory footprint analysis reveals that our routing mechanism introduces negligible overhead to the system. As detailed in Table \ref{tab:running_times}, the SecureRouter consumes 38.20 MB of memory, a marginal increase of only 1.12 MB (approximately 3\%) over the standalone BERT\_tiny expert (38.00 MB). This confirms that the overhead imposed by the routing logic and model management is computationally lightweight. Furthermore, when contrasted with the memory demands of standard dense models—1054.00 MB for BERT\_base and 3114.00 MB for BERT\_large—our approach maintains a memory profile comparable to the smallest expert in the pool. This demonstrates that SecureRouter achieves dynamic inference acceleration without incurring the prohibitive memory costs typically associated with deploying larger or ensemble-based architectures in secure inference environments.

\section{Conclusion}

In this paper, first, we present SecureRoute, an end-to-end encrypted routing and inference framework designed to enable efficient and accurate MPC-based Transformer instead of using fixed model. Secondly, we detail the online inference protocol, where two servers to use secret-shared input embeddings from the client to privately select and execute an optimal model. Lastly, in the offline process, we introduce an MPC cost aware secure router training algorithm to predict inference cost and utility from encrypted inputs. Extensive experiments on the GLUE benchmark demonstrate that SecureRoute achieves significant speedups over static baselines without compromising model accuracy. Ultimately, this work establishes a scalable path for deploying large-scale, privacy-preserving AI services in latency-critical applications.

\newpage

\bibliographystyle{ACM-Reference-Format}
\bibliography{bib}

\appendix

\section{Expert Pool Scalability}

We investigate the impact of expert pool size $K$ on routing quality. Starting from the default 3-expert pool (Tiny, Base, Large), we construct pools of $K \in \{2, 3, 4, 5\}$ experts by adding intermediate BERT variants (Small, Mini). All configurations use the same router architecture, training hyperparameters, and cost-aware loss weights ($\alpha{=}0.05$, $\beta{=}0.08$) on the MRPC task to ensure a controlled comparison.

\begin{table}[h]
  \centering
  \resizebox{\columnwidth}{!}{%
    \begin{tabular}{clccc}
      \toprule
      \textbf{K} & \textbf{Expert Pool} & \textbf{Costs} & \textbf{F1} & \textbf{Acc.} \\
      \midrule
      2 & Tiny, Large & [2, 13] & 89.75 & 85.05 \\
      3 & Tiny, Base, Large & [2, 7, 13] & \textbf{90.14} & \textbf{85.78} \\
      4 & Tiny, Small, Base, Large & [2, 4, 7, 13] & 89.23 & 84.31 \\
      5 & Tiny, Mini, Small, Base, Large & [2, 3, 4, 7, 13] & 89.19 & 84.31 \\
      \bottomrule
    \end{tabular}%
  }
  \caption{Expert pool scalability on MRPC. $K{=}3$ achieves the best accuracy--efficiency balance; adding more experts introduces routing ambiguity without improving quality.}
  \label{tab:scalability}
\end{table}

As shown in Table~\ref{tab:scalability}, the 3-expert pool ($K{=}3$) achieves the best F1 of 90.14. The 2-expert pool ($K{=}2$) lacks a mid-range expert, forcing the router to choose between two extremes and reducing F1 by 0.39. Adding a fourth or fifth expert ($K{=}4, 5$) does not improve quality; the additional mid-range models overlap in capacity, increasing routing ambiguity without providing complementary coverage. This confirms that a small, well-separated expert pool is sufficient for effective cost-aware routing.

\section{Cost Sensitivity Analysis}

To validate that the router's behavior is genuinely driven by the cost-aware loss $L_{\text{cost}}$ rather than memorized heuristics, we vary the expert cost vector while keeping all other training conditions fixed (MRPC, $K{=}3$, $\alpha{=}0.05$, $\beta{=}0.08$).

\begin{table}[h]
  \centering
  \resizebox{\columnwidth}{!}{%
    \begin{tabular}{lccccc}
      \toprule
      \textbf{Cost Profile} & \textbf{Costs} & \textbf{F1} & \textbf{Acc.} & \textbf{\makecell{Tiny\\Route\%}} & \textbf{\makecell{Large\\Route\%}} \\
      \midrule
      Baseline & [2, 7, 13] & 89.70 & 84.80 & 15.9 & 61.8 \\
      Scale$\times$0.7 & [1.4, 4.9, 9.1] & 90.53 & 86.52 & 11.3 & 50.7 \\
      Scale$\times$1.5 & [3, 10.5, 19.5] & 90.34 & 86.27 & 13.2 & 48.0 \\
      Flat & [5.1, 7.3, 9.5] & 90.88 & 87.01 & 6.6 & 63.5 \\
      Steep & [1, 7, 19.5] & 90.10 & 85.78 & 13.0 & 45.8 \\
      Reversed & [13, 7, 2] & \textbf{91.48} & \textbf{87.99} & 2.7 & 90.4 \\
      \bottomrule
    \end{tabular}%
  }
  \caption{Cost sensitivity analysis on MRPC. The router adapts its expert selection in response to different cost vectors, confirming that $L_{\text{cost}}$ drives routing behavior.}
  \label{tab:cost_sensitivity}
\end{table}

Table~\ref{tab:cost_sensitivity} reveals clear cost-responsive routing behavior. Under the \emph{Steep} profile, where the Large expert is 19.5$\times$ more expensive than Tiny, the router aggressively reduces Large usage to 45.8\%, compared to 61.8\% under the Baseline profile. Under the \emph{Flat} profile, where all experts have similar costs, the cost penalty becomes negligible and the router defaults to accuracy-maximizing selections (F1 = 90.88). The \emph{Reversed} configuration---where Tiny is the most expensive and Large is the cheapest---causes the router to route 90.4\% of queries to Large, yielding the highest F1 of 91.48. These results confirm that the cost-aware loss effectively steers routing decisions according to the provided cost structure, rather than relying on fixed architectural biases.

\end{document}